\documentclass[twocolumn,floatfix,pre,aps,showpacs]{revtex4}
\usepackage{graphicx}
\usepackage{amsmath}

\begin{document}

\title{Desynchronized wave patterns in synchronized chaotic regions
of coupled map lattices}

\author{P.~Palaniyandi}
\affiliation{Centre for Nonlinear Dynamics, Department of Physics,
Bharathidasan University, Tiruchirapalli 620 024, India}

\author{P.~Muruganandam}
\affiliation{Centre for Nonlinear Dynamics, Department of Physics,
Bharathidasan University, Tiruchirapalli 620 024, India}

\author{M.~Lakshmanan}
\affiliation{Centre for Nonlinear Dynamics, Department of Physics,
Bharathidasan University, Tiruchirapalli 620 024, India}

\date{\today}

\begin{abstract}

We analyze the size limits of coupled map lattices with diffusive
coupling at the crossover of low-dimensional to high-dimensional chaos.
We investigate the existence of standing-wave-type periodic patterns,
within the low-dimensional limit, in addition to the stable synchronous
chaotic states depending upon the initial conditions. Further, we bring
out a controlling mechanism to explain the emergence of standing-wave
patterns in the coupled map lattices. Finally, we give an analytic
expression in terms of the unstable periodic orbits of the isolated map
to represent the standing-wave patterns.

\end{abstract}


\maketitle

\section{Introduction}
\label{sec1}

Coupled dynamical systems often arise in nature whenever a collective
or cooperative phenomenon is favoured~\cite{lakshman_murali:book:96,
Pikovsky_etal:book:01, lakshmanan:book:03:01, muruganandam:99:01}.  In
particular, the coupled map lattice with diffusive coupling (CML)
provides a prototype model to study various features associated with
the cooperative evolution of constituent systems~\cite{kaneko:86:01,
kaneko:92:01, lind_etal:02:01, gerson_anand:03-01,kaneko:00:01}.  One
of the important properties of such CMLs is that they exhibit size
instability, that is, there is a critical size on the number of
constituents for which stable synchronous chaotic state exists.
Increasing the number of constituents beyond this limit leads to the
occurrence of spatially incoherent behaviour (eg., high-dimensional
chaos). For example, Bohr and Christensen~\cite{bohr:89:01} have
studied such size instability behaviour in a two-dimensional coupled
logistic lattice. Similar desynchronization has been found in arrays of
coupled systems represented by nonlinear
oscillators~\cite{muruganandam:99:01,
heagy:94:01,pecora:98:01,restrepo:04:01}. The stability of synchronous
chaos in coupled dynamical systems plays an important role in the study
of pattern formation, spatiotemporal chaos,
etc.~\cite{muruganandam:99:01, heagy:94:01, bohr:89:01,
rangarajan:02:01, chen:03:01}.

In general, these studies on size instability are valid in most
situations. However, we have noted that in certain circumstances there
is an ambiguity in dealing with these systems below the critical system
sizes. To be specific, there exist certain nontrivial ranges of initial
conditions for which the CML admits spatial and temporally periodic 
solutions in contrast to the usually expected stable synchronous chaos.
In this brief report, we show numerically the coexistence of such
periodic states with the stable synchronous chaotic state well below
the critical system size and explain the underlying mechanism.

\section{Size instability in Coupled Map Lattices with diffusive
coupling} \label{sec2}

Consider an one-dimensional coupled map  lattice with nearest-neighbour
diffusive coupling~\cite{kaneko:86:01,kaneko:92:01}
\begin{align}
x_{n+1}^j =   f(x_n^j) + \epsilon  \left[ f(x_n^{j-1}) +
f(x_n^{j+1}) - 2f(x_n^j) \right], \label{cml}
\end{align}
where $j$ $(=0,1,2,\cdots, L-1)$ represents the lattice sites and
$L$ is the system size, subject to periodic boundary conditions.

The stability analysis of the synchronized chaotic state defined
by $s_n = x^0 = x^1 = \cdots = x^{L-1}$ in the above CML using the
procedure derived originally for coupled oscillators by Heagy
\emph{et al}~\cite{heagy:94:01}, gives the relation connecting
transverse Lyapunov exponents (TLEs), in terms of the Lyapunov
exponent of single (isolated) map, $\lambda^0$ as
\begin{align}
\lambda^k=\lambda^0+\ln\left[1-4\epsilon \text{sin}^2\left(\frac
{\pi k}{L} \right)\right]. \label{tr_lya}
\end{align}
The synchronous state is stable only if the TLEs ($\lambda^k$,
$k=1,2,\cdots\,L-1$) are all negative.

The above relation (\ref{tr_lya}) can also be obtained by means
of a direct perturbation of the form
\begin{align}
x_n^j = s_n + \delta\, \exp\left(i\frac{2\pi k}{L}\right)\,
\exp\left(\lambda^k n\right), \;\; \delta \ll 1,
\end{align}
as considered by Bohr and Christensen~\cite{bohr:89:01}.

The synchronous state loses its stability when the long waves
(lowest mode) are unstable~\cite{bohr:89:01}. This means that for
$\lambda^1 > 0$ the synchronous state is unstable. Thus,
substituting $\lambda^1 = 0$ in Eq.~(\ref{tr_lya}), one obtains
the maximum/critical lattice size $(L_c)$ that supports stable
synchronous state as
\begin{align}
L_c=\text{int}\left( \frac{\pi}{\text{sin}^{-1} \left(
\sqrt{\frac{1-e^{-\lambda_0}}{4\epsilon}} \right)} \right).
\label{lmax}
\end{align}
Now, let us consider a coupled logistic lattice with diffusive coupling 
(CLL) where each lattice site in Eq.~(\ref{cml}) is occupied by the 
logistic map
\begin{align}
f(x) = \mu x(1-x), \;\; x \in (0,1), \;\; \mu \in (0,4).
\label{fx:logi}
\end{align}
In particular for the choice $\mu = 3.5732$, the  Lyapunov exponent of
single (isolated) map is positive (i.e., $\lambda^0 \sim 0.057 > 0$)
and chaotic. In this case, for coupling strength $\epsilon =0.2$, the
critical lattice size ($L_c$)  is found to be $11$ from
Eq.~(\ref{lmax}).  That is,  upto the lattice size $11$ the CLL
exhibits  synchronous chaos and for lattice size $\ge 12$ the
synchronization is found to be lost, thereby confirming  the size
instability in the diffusively coupled logistic lattices.

\section{Existence of Multiple stable states in Coupled Map Lattices}
\label{sec3} 

When random initial conditions are assumed, in most cases, the above CLL
with diffusive coupling exhibits stable synchronous chaos for $2 \le L \le 11$ as
predicted by Eq.~(\ref{lmax}). However, there are certain ranges
of initial conditions for which CLL shows some interesting
asynchronous spatiotemporal patterns even for $L < L_c$. For
\begin{figure}[!ht]
\centering{\includegraphics[width=0.75\linewidth]{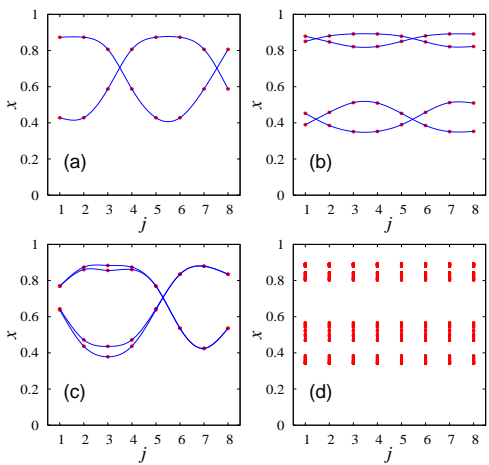}} 
\caption{The possible spatiotemporal patterns in coupled logistic lattice
(\ref{cml}) and (\ref{fx:logi}) with $L=8$ for different initial
conditions. (a) single standing wave, (b) double standing waves
(c) single standing wave with a temporal period-2 orbit in one
half and period-4 orbit in the other half of the lattice
 and (d) synchronized chaos.}
\label{n8dyn_n}
\end{figure}
example, for $L = 8$ and with the choice of initial conditions
$\{x^j_0\}_{j=0}^{L-1} = $ $\{0.1$, $0.01$, $0.7$, $0.2$,  $0.65$,
$0.1$, $0.15$, $0.001\}$ or several nearby initial conditions, the
CLL exhibits a standing wave type pattern as shown in
Fig.~\ref{n8dyn_n}(a). But for the same lattice size, if we choose
a different set of initial conditions, $\{x^j_0\}_{j=0}^{L-1} = $
$\{0.0004$, $0.0001$, $0.0003$, $0.0005$, $0.00045$, $0.00018$,
$0.00016$, $0.00001\}$ or the nearby points, two standing waves
with different amplitudes are produced within the CLL as in
Fig.~\ref{n8dyn_n}(b). Similarly a disturbed standing wave pattern
as shown in Fig.~\ref{n8dyn_n}(c) is possible to be exhibited by the
CLL for many choices of initial conditions. A synchronous
chaos, as one would expect for $L = 8$ from the theory, is also
exhibited by the CLL as depicted in Fig.~\ref{n8dyn_n}(d), for
most of the random choices of initial conditions. This kind of
multiple stable solutions is also observed in  the CLL for other
lattice sizes, namely $L=6$, $7$, $9$ and $10$ which are also
less than $L_c$. The occurrence of various spatiotemporal patterns
for different lattice sizes of the CLL and their percentage of
occurrence are shown in Table~\ref{tab_pattern}. In order to
quantify the percentage of occurrence of different spatiotemporal
patterns, we have used $10^6$ sets of random initial conditions
(i.c.'s) in the interval $0$ to $1$ and identified the number of
i.c.'s which lead to a specific pattern as indicated in
Table~\ref{tab_pattern}. We have further confirmed our assertion
by analyzing the same systems in different computing environments
such as Intel Pentium 4, Sun Sparc server/workstation and Compaq
Alpha workstation.

In addition, by considering the whole CML of size $L$ as a single
$L$-dimensional map we have verified that the above spatiotemporal
periodic structures are essentially the stable fixed points of this
map. For example, the periodic structure in Fig.~\ref{n8dyn_n}(a)
represents a fixed of point of period-2 of the eight-dimensional map
and the eigenvalues of the corresponding Jacobian matrix all are having
magnitude less than unity. In a similar fashion one can verify that all
the periodic structures are the stable fixed points of corresponding
periods. Thus, in addition to the stable synchronized manifold, there
exists other invariant sets corresponding to stable periodic
structures.

\begin{table}[!ht]
\caption{The possible spatiotemporal patterns for different
lattice sizes and their percentage of occurrence for the coupled
logistic lattice sampled over a set of $10^{6}$ random initial
conditions (IC's). The parameters are fixed as $\mu = 3.5732$ in
Eq.~(\ref{fx:logi}) and $\epsilon = 0.2$ in Eq.~(\ref{cml}). Here,
SW: standing wave, sync.: synchronized.} \centering
\begin{tabular}{clr|clr}
\hline \multicolumn{1}{c}{size~~~}
    & \multicolumn{1}{c}{spatiotemporal~~~}
    & \multicolumn{1}{c|}{\% of}
    & \multicolumn{1}{c}{~size~~~}
    & \multicolumn{1}{c}{spatiotemporal~~~}
    & \multicolumn{1}{c}{\% of} \\
\multicolumn{1}{c}{($L$)~~~}
    & \multicolumn{1}{c}{patterns}
    & \multicolumn{1}{c|}{IC's}
    & \multicolumn{1}{c}{($L$)~~~}
    & \multicolumn{1}{c}{patterns}
    & \multicolumn{1}{c}{IC's} \\
\hline
 $6$ & single SW      & $ 9$ & $10$ & double SWs   & $43$ \\
     & sync. chaos    & $91$ &      & sync. chaos  & $45$ \\
 $7$ & single SW      & $22$ &      & others       & $12$ \\
     & sync. chaos    & $78$ & $11$ & double SWs   & $48$ \\
 $8$ & single SW      & $21$ &      & four SWs     & $ 2$ \\
     & double SWs     & $10$ &      & sync. chaos  & $34$ \\
     & sync. chaos    & $66$ &      & others       & $16$ \\
     & others         & $ 3$ & $12$ & double SWs   & $24$  \\
 $9$ & double SWs     & $38$ &      & four SWs    & $28$ \\
     & sync. chaos    & $54$ &      & sync. chaos  & $0.2$ \\
     & others         & $ 8$ &      & others       & $47.8$ \\
\hline
\end{tabular}
\label{tab_pattern}
\end{table}

\section{Emergence of standing wave patterns by controlling}
\label{sec4}

The extraordinary behaviour of the CML with diffusive coupling showing
standing wave patterns well below the critical lattice size ($L_c$) can
be explained as follows. The second term in the right hand side of
Eq.~(\ref{cml}), can be considered as a kind of force or perturbation
applied to every lattice point in the CML and we call it as the
coupling force. In fact this force on a particular lattice point is
developed either due to a mismatch in the parameters of the
neighbouring lattice points or due to differences in their initial
conditions or both. If the neighbouring lattice points are identical
then this force is formed due to variation in the initial conditions
and our system indeed falls under this category. In general, the
strength of the coupling force at all the lattice points approaches
zero when they oscillate towards synchronization with their neighbours,
and usually this will happen for $L \le L_c$. But in the case of
$L>L_c$,  this force at every lattice point oscillates periodically or
in a chaotic manner, giving rise to various spatiotemporal patterns,
including standing waves. However, as we have pointed out above that
under certain circumstances, (i.e., for certain ranges of initial
conditions), even for $L < L_c$, the coupling force of each and every
lattice point oscillates periodically with different amplitudes and
thereby makes the CML to exhibit spatiotemporal periodic (standing
wave) solutions. In general, the periodic oscillations in the coupling
force may be of any period and this fixes the number of standing waves
produced within the lattice. 

In order to understand the mechanism behind the emergence of
spatiotemporal periodic structure, let us now consider a specific
case of the periodically oscillating coupling force of period two,
that is, the force oscillating between two fixed amplitudes, say,
$k_1$ and $k_2$ so that a single standing wave is formed in the
CML. Then the amplitudes of the coupling force at the $j^{th}$
lattice site will alternate between the numerical values $k^{j}_1$
and $k^{j}_2$, where $j=1,2,\ldots,L$.  Thus, the evolution of
$j^{th}$ map in the CML can be effectively described by the
equation
\begin{align}
& x^{j}_{n+1}  \, = f(x^{j}_{n}) + k^{j}_1, \;\; 
& x^{j}_{n+2}  \, = f(x^{j}_{n+1}) + k^{j}_2. \label{cml_dc}
\end{align}
For a given lattice point $j$, this is nothing but a single
logistic map with a periodic kick of period-2 (modified map). It
is now obvious to note from Eqs.~(\ref{cml_dc}) and (\ref{cml})
that the original coupled map lattice exhibiting a standing wave
pattern can be decomposed into $L$ number of modified maps such
that the dynamics of (\ref{cml}) is essentially mimicked by the
set (\ref{cml_dc}). 
Thus studying the evolution of $L$ decoupled modified maps
(\ref{cml_dc}) with allowed sets of values for $k^{j}_1$ and
$k^{j}_2$  is equivalent to that of the original CML, given by
Eq.~(\ref{cml}).

In general, a chaotically evolving system can be controlled to a
stable periodic orbit by the addition of an appropriate constant
or periodic external bias~\cite{pyragas:92:01,
loskuto_chaos:94:01, lakshmanan:book:03:01, venkatesan:03:01,
palaniyandi:04:01}. As a consequence, one can expect a stable
fixed point solution for the modified map (\ref{cml_dc}) for
appropriate forcing amplitudes ($k_1$ and $k_2$), with same set
of parameters for which the original map, $x^{j}_{n+1}=
f(x^{j}_{n})$, exhibits chaotic solution. Thus there exists a
possibility for obtaining periodic solutions of period two for
the constituent map within the CML even though its
parameter is in the chaotic region of the individual map. 

\begin{figure}[!ht]
\centering{\includegraphics[width=0.67\linewidth]{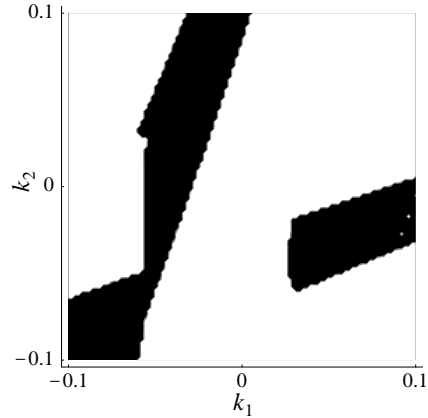}}
\caption{The phase diagram in the $k_1-k_2$ plane for the modified
map (\ref{cml_dc}) with $f(x) = \mu x (1-x)$, $\mu = 3.5732$, showing
the regions which correspond to the periodic solution of period
$2$ (dark region).} \label{phase}
\end{figure}

In the case of coupled logistic lattice with parameters
mentioned in the previous section, the allowed region of
forcing amplitudes which lead the decoupled modified map
(\ref{cml_dc}) to exhibit period two solution is shown in
Fig.~\ref{phase}.  For $L=8$, we have calculated the
amplitudes of the coupling force [that is, the second term in
the right hand side of Eq.~(\ref{cml})] of period-$2$ for the
initial conditions, $\{x^j_0\}_{j=0}^{L-1} = $ $\{0.1$,
$0.01$, $0.7$, $0.2$, $0.65$, $0.1$, $0.15$, $0.001\}$, and
these are shown in Table~\ref{table1}. Now one can easily
check that these amplitudes of the coupling force fall in the
region specified by the phase diagram shown in
Fig.~\ref{phase}.  Also, the shift invariant property of the
CLL ensures that there is no temporal variation if we shift the initial
conditions of each lattice point in the CML to its neighbour spatially. In this
case, the wave pattern will also make only a corresponding shift.  We have made
similar investigations for higher periodic standing waves which lead to same
type of conclusions based on the appropriate periodic nature of the coupling
force. 
\begin{table}[!ht]
\centering \caption{Amplitudes of the coupling force $k_1$ and
$k_2$ in the coupled map lattices for $L=8$ with period-$2$ for
initial conditions, $\{x^j_0\}_{j=0}^{L-1} = $ $\{0.1$, $0.01$,
$0.7$, $0.2$,  $0.65$, $0.1$, $0.15$, $0.001\}$.}
\begin{tabular}{crr|crr}
\hline \multicolumn{1}{c}{lattice}
    & \multicolumn{1}{c}{$k_1^j$}
    & \multicolumn{1}{c|}{$k_2^j$}
    & \multicolumn{1}{c}{lattice}
    & \multicolumn{1}{c}{$k_1^j$}
    & \multicolumn{1}{c}{$k_2^j$} \\
\multicolumn{1}{c}{site(j)}
    & \multicolumn{1}{c}{($\times 10^{-2}$)}
    & \multicolumn{1}{c|}{($\times 10^{-2}$)}
    & \multicolumn{1}{c}{site(j)}
    & \multicolumn{1}{c}{($\times 10^{-2}$)}
    & \multicolumn{1}{c}{($\times 10^{-2}$)} \\
\hline
$1$ & $-0.17626$   & $ 3.24717$ & $5$ & $ 3.24717$   & $-0.17626$  \\
$2$ & $-5.98282$   & $ 2.91190$ & $6$ & $ 2.91190$   & $-5.98282$ \\
$3$ & $ 2.91190$   & $-5.98282$ & $7$ & $-5.98282$   & $ 2.91190$ \\
$4$ & $ 3.24717$   & $-0.17626$ & $8$ & $-0.17626$   & $ 3.24717$ \\
\hline
\end{tabular}
\label{table1}
\end{table}

So, if it is possible to control the coupling force  to fall in
the region which corresponds to a periodic solution, then one can
obtain standing wave patterns irrespective of the size of the
lattices.  This is in fact possible by choosing appropriate
initial conditions to each lattice point and this explains the
occurrence of standing waves (asynchronous) as shown in
Figs.~\ref{n8dyn_n}(a) and \ref{n8dyn_n}(b) in the CLL well below
the critical lattice size ($L_c$). Similar explanation holds good
for Fig.~\ref{n8dyn_n}(c). The same principle is involved in the
occurrence of standing waves even for $L>L_c$~\cite{kaneko:92:01}.

\section{Analytical expression for standing wave patterns}

\label{sec5}

From a careful numerical analysis, we have observed that the
nodes and antinodes of standing waves are formed at or very close to the UPOs
\begin{figure}[!ht]
\centering{\includegraphics[width=0.75\linewidth]{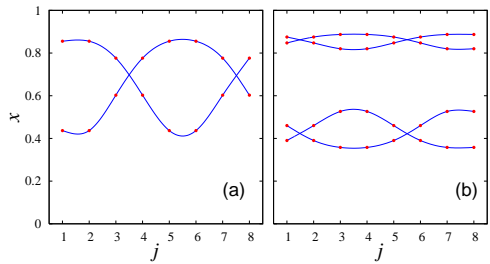}}
\caption{Standing wave patterns obtained from Eq.~(\ref{swave})
for $L=8$: (a) standing wave with period-1 UPO at nodes and
period-2 UPO at antinodes and (b) period-2 UPO at nodes and
period-4 UPO at antinodes. } \label{n8dyn_a}
\end{figure}
of the isolated logistic map.  Keeping this in mind, we propose an expression
for the standing wave pattern~\cite{main:93:01} of the form
\begin{align}\label{swave}
x^j = u_k + A_k \sin\left(\frac{m\pi (j-\delta)}{L}\right)
\cos\left(\pi i\right),
\end{align}
where,
\begin{align}
A_k=  \left\{
\begin{array}{ll}
A^{\mbox{\scriptsize max}}_k, & \text{~~~if } \sin\left(\frac{m\pi
(j-\delta)}{L}\right)
\cos\left(\pi i\right) > 0   \nonumber \\
A^{\mbox{\scriptsize min}}_k, & \text{~~~if } \sin\left(\frac{m\pi
(j-\delta)}{L}\right) \cos\left(\pi i\right) < 0
\end{array}
\right.
\end{align}
where the discrete index $j=0,1,2,\cdots, L-1$ corresponds to the lattice site,
$m$ denotes the mode of the waves, and $k$ and $i$ represent the nodes and
antinodes of different standing waves present in the pattern, which can take
values from $1$ to $p$ and $1$ to $2p$, respectively and $p$ is the period of
UPO. Also in the above equation (\ref{swave}), $u_k$'s represent the values of
the UPOs of the isolated logistic map at the node of the $k$-th standing wave,
and $A^{\mbox{\scriptsize max}}_k$ and $A^{\mbox{\scriptsize min}}_k$ are the absolute values of the differences
between UPOs at the node and UPOs at high and low amplitudes at the antinodes
of the $k$th standing waves, respectively.
The wave patterns with one and two number of standing waves obtained using
Eq.~(\ref{swave}) for $L=8$ are shown in Fig.~\ref{n8dyn_a}, whereas for the
same lattice size, the numerically obtained pattern which have been discussed
in Sec.~\ref{sec3} is shown in Figs.~\ref{n8dyn_n}(a) and \ref{n8dyn_n}(b).
One observes that these two figures coincide very closely.  

\section{Summary and Conclusion}
\label{sec6}

In this report, we have pointed out that coupled map lattices with
diffusice coupling exhibit multiple stable states for the same set of
parameters with respect to the initial conditions. It has also been
shown that by choosing appropriate initial conditions, one can obtain
different standing wave type patterns for such coupled map lattices
even for lattice size much less than the critical system size $L_c$,
where one normally would expect synchronized chaos. In addition, we
have proposed the mechanism behind the occurrence  of such standing
wave patterns.

\acknowledgments

This work has been supported by the National Board for Higher
Mathematics, Department of Atomic Energy, Government of India and
the Department of Science and Technology, Government of India
through research projects.

\end{document}